\documentclass[superscriptaddress,floatfix,twocolumn,showpacs,preprintnumbers,amsmath,amssymb]{revtex4}

\usepackage{graphicx}
\usepackage{dcolumn}
\usepackage{bm}

\begin{document}


\title{Measurements of $\bm{G_{E}^{n}/G_{M}^{n}}$ from the
$\bm{^{2}}$H($\vec{\bm{e}}\bm{,e'}\vec{\bm{n}}$)$\bm{^{1}}$H Reaction to 
$\bm{Q^{2}=1.45}$ (GeV/$\bm{c}$)$\bm{^{2}}$}

\author{R.~Madey}
\affiliation{Kent State University, Kent, Ohio 44242}
\affiliation{Thomas Jefferson National Accelerator Facility, Newport News,
Virginia 23606}
\author{A.~Yu.~Semenov}
\affiliation{Kent State University, Kent, Ohio 44242}
\author{S.~Taylor}
\affiliation{Massachusetts Institute of Technology, Cambridge, Massachusetts 02139}
\author{B.~Plaster}
\affiliation{Massachusetts Institute of Technology, Cambridge, Massachusetts 02139}
\author{A.~Aghalaryan}
\affiliation{Yerevan Physics Institute, Yerevan 375036, Armenia}
\author{E.~Crouse}
\affiliation{The College of William and Mary, Williamsburg, Virginia 23187}
\author{G.~MacLachlan}
\affiliation{Ohio University, Athens, Ohio 45701}
\author{S.~Tajima}
\affiliation{Duke University and TUNL, Durham, North Carolina 27708}
\author{W.~Tireman}
\affiliation{Kent State University, Kent, Ohio 44242}
\author{Chenyu Yan}
\affiliation{Kent State University, Kent, Ohio 44242}
\author{A.~Ahmidouch}
\affiliation{North Carolina A\&T State University, Greensboro, North Carolina 27411}
\author{B.~D.~Anderson}
\affiliation{Kent State University, Kent, Ohio 44242}
\author{H.~Arenh\"{o}vel}
\affiliation{Johannes Gutenberg-Universit\"{a}t, D-55099 Mainz, Germany}
\author{R.~Asaturyan}
\affiliation{Yerevan Physics Institute, Yerevan 375036, Armenia}
\author{O.~Baker}
\affiliation{Hampton University, Hampton, Virginia, 23668}
\author{A.~R.~Baldwin}
\affiliation{Kent State University, Kent, Ohio 44242}
\author{D.~Barkhuff}
\altaffiliation[Present Address: ]{Renaissance Technologies, 
 600 Route 25A, 
 East Setauket, New York 11733.}
\affiliation{Massachusetts Institute of Technology, Cambridge, Massachusetts 02139}
\author{H.~Breuer}
\affiliation{University of Maryland, College Park, Maryland 20742}
\author{R.~Carlini}
\affiliation{Thomas Jefferson National Accelerator Facility, Newport News,
Virginia 23606}
\author{E.~Christy}
\affiliation{Hampton University, Hampton, Virginia, 23668}
\author{S.~Churchwell}
\altaffiliation[Present Address: ]{Dept. of Physics and Astronomy, 
University of Canterbury, 
Private Bag 4800, 
Christchurch 8020, New Zealand.}
\affiliation{Duke University and TUNL, Durham, North Carolina 27708}
\author{L.~Cole}
\affiliation{Hampton University, Hampton, Virginia, 23668}
\author{S.~Danagoulian}
\affiliation{Thomas Jefferson National Accelerator Facility, Newport News,
Virginia 23606}
\affiliation{North Carolina A\&T State University, Greensboro, North Carolina 27411}
\author{D.~Day}
\affiliation{University of Virginia, Charlottesville, Virginia 22904}
\author{T.~Eden}
\altaffiliation[Present Address: ]{Atmospheric Chemistry Division, National Center 
for Atmospheric Research, Boulder, Colorado 80307.}
\affiliation{Kent State University, Kent, Ohio 44242}
\affiliation{Hampton University, Hampton, Virginia, 23668}
\author{M.~Elaasar}
\affiliation{Southern University at New Orleans, New Orleans, Louisiana 70126}
\author{R.~Ent}
\affiliation{Thomas Jefferson National Accelerator Facility, Newport News,
Virginia 23606}
\author{M.~Farkhondeh}
\affiliation{Massachusetts Institute of Technology, Cambridge, Massachusetts 02139}
\author{H.~Fenker}
\affiliation{Thomas Jefferson National Accelerator Facility, Newport News,
Virginia 23606}
\author{J.~M.~Finn}
\affiliation{The College of William and Mary, Williamsburg, Virginia 23187}
\author{L.~Gan}
\affiliation{Hampton University, Hampton, Virginia, 23668}
\author{K.~Garrow}
\affiliation{Thomas Jefferson National Accelerator Facility, Newport News,
Virginia 23606}
\author{P.~Gueye}
\affiliation{Hampton University, Hampton, Virginia, 23668}
\author{C.~R.~Howell}
\affiliation{Duke University and TUNL, Durham, North Carolina 27708}
\author{B.~Hu}
\affiliation{Hampton University, Hampton, Virginia, 23668}
\author{M.~K.~Jones}
\affiliation{Thomas Jefferson National Accelerator Facility, Newport News,
Virginia 23606}
\author{J.~J.~Kelly}
\affiliation{University of Maryland, College Park, Maryland 20742}
\author{C.~Keppel}
\affiliation{Hampton University, Hampton, Virginia, 23668}
\author{M.~Khandaker}
\affiliation{Norfolk State University, Norfolk, Virginia 23504}
\author{W.-Y.~Kim}
\affiliation{Kyungpook National University, Taegu 702-701, Korea}
\author{S.~Kowalski}
\affiliation{Massachusetts Institute of Technology, Cambridge, Massachusetts 02139}
\author{A.~Lai}
\affiliation{Kent State University, Kent, Ohio 44242}
\author{A.~Lung}
\affiliation{Thomas Jefferson National Accelerator Facility, Newport News,
Virginia 23606}
\author{D.~Mack}
\affiliation{Thomas Jefferson National Accelerator Facility, Newport News,
Virginia 23606}
\author{D.~M.~Manley}
\affiliation{Kent State University, Kent, Ohio 44242}
\author{P.~Markowitz}
\affiliation{Florida International University, Miami, Florida 33199}
\author{J.~Mitchell}
\affiliation{Thomas Jefferson National Accelerator Facility, Newport News,
Virginia 23606}
\author{H.~Mkrtchyan}
\affiliation{Yerevan Physics Institute, Yerevan 375036, Armenia}
\author{A.~K.~Opper}
\affiliation{Ohio University, Athens, Ohio 45701}
\author{C.~Perdrisat}
\affiliation{The College of William and Mary, Williamsburg, Virginia 23187}
\author{V.~Punjabi}
\affiliation{Norfolk State University, Norfolk, Virginia 23504}
\author{B.~Raue}
\affiliation{Florida International University, Miami, Florida 33199}
\author{T.~Reichelt}
\affiliation{Rheinische Friedrich-Wilhelms-Universit\"{a}t, D-53115 Bonn,
Germany}
\author{J.~Reinhold}
\affiliation{Florida International University, Miami, Florida 33199}
\author{J.~Roche}
\affiliation{The College of William and Mary, Williamsburg, Virginia 23187}
\author{Y.~Sato}
\affiliation{Hampton University, Hampton, Virginia, 23668}
\author{N.~Savvinov}
\affiliation{University of Maryland, College Park, Maryland 20742}
\author{I.~A.~Semenova}
\affiliation{Kent State University, Kent, Ohio 44242}
\author{W.~Seo}
\affiliation{Kyungpook National University, Taegu 702-701, Korea}
\author{N.~Simicevic}
\affiliation{Louisiana Tech University, Ruston, Louisiana 71272}
\author{G.~Smith}
\affiliation{Thomas Jefferson National Accelerator Facility, Newport News,
Virginia 23606}
\author{S.~Stepanyan}
\affiliation{Yerevan Physics Institute, Yerevan 375036, Armenia}
\affiliation{Kyungpook National University, Taegu 702-701, Korea}
\author{V.~Tadevosyan}
\affiliation{Yerevan Physics Institute, Yerevan 375036, Armenia}
\author{L.~Tang}
\affiliation{Hampton University, Hampton, Virginia, 23668}
\author{P.~E.~Ulmer}
\affiliation{Old Dominion University, Norfolk, Virginia 23529}
\author{W.~Vulcan}
\affiliation{Thomas Jefferson National Accelerator Facility, Newport News,
Virginia 23606}
\author{J.~W.~Watson}
\affiliation{Kent State University, Kent, Ohio 44242}
\author{S.~Wells}
\affiliation{Louisiana Tech University, Ruston, Louisiana 71272}
\author{F.~Wesselmann}
\affiliation{University of Virginia, Charlottesville, Virginia 22904}
\author{S.~Wood}
\affiliation{Thomas Jefferson National Accelerator Facility, Newport News,
Virginia 23606}
\author{Chen Yan}
\affiliation{Thomas Jefferson National Accelerator Facility, Newport News,
Virginia 23606}
\author{S.~Yang}
\affiliation{Kyungpook National University, Taegu 702-701, Korea}
\author{L.~Yuan}
\affiliation{Hampton University, Hampton, Virginia, 23668}
\author{W.-M.~Zhang}
\affiliation{Kent State University, Kent, Ohio 44242}
\author{H.~Zhu}
\affiliation{University of Virginia, Charlottesville, Virginia 22904}
\author{X.~Zhu}
\affiliation{Hampton University, Hampton, Virginia, 23668}

\collaboration{The Jefferson Laboratory E93-038 Collaboration}

\date{\today}

\begin{abstract}
We report new measurements of the ratio of the electric form factor to
the magnetic form factor of the neutron, $G_{E}^{n}/G_{M}^{n}$,
obtained via recoil polarimetry from the quasielastic
$^{2}$H$(\vec{e},e'\vec{n})^{1}$H reaction at $Q^2$ values of 0.45, 
1.13, and 1.45 (GeV/$c$)$^{2}$ with relative statistical uncertainties
of 7.6 and 8.4\% at the two higher $Q^{2}$ points, which were not reached
previously via polarization measurements. Scale and systematic uncertainties 
are small.
\end{abstract}


\maketitle

The nucleon elastic electromagnetic form factors are fundamental
quantities needed for an understanding of nucleon and nuclear
structure.  The evolution of the electric and magnetic form factors
with $Q^{2}$, the square of the four-momentum transfer, is related to
the charge and current distributions within the nucleon.  Precision
measurements of the electromagnetic form factors are important for
tests of non-perturbative quantum chromodynamics (QCD) either on the
lattice or in models.  Measurements of the neutron electric form
factor, $G_{E}^{n}$, have been impeded by the lack of a free neutron
target and the small value of $G_{E}^{n}$ relative to the neutron
magnetic form factor, $G_{M}^{n}$; however, with the advent of high
duty-factor polarized electron beam facilities, experiments employing
recoil polarimeters \cite{eden94,herberg99}, polarized
$^{3}$He targets \cite{meyerhoff94,rohe99,golak01}, and
polarized deuterium targets \cite{passchier99,zhu01} have yielded the
first precision measurements of $G_{E}^{n}$.  These polarization
measurements of $G_{E}^{n}$ are limited to $Q^{2}~{\leq}~0.67$
(GeV/$c$)$^{2}$ and are, within errors, consistent with the Galster
parameterization \cite{galster71}.  In addition to polarization
measurements, $G_{E}^{n}$ has been extracted from a theoretical
analysis of the deuteron quadrupole form factor
\cite{schiavilla01} for $Q^2$ values up to 1.6 (GeV/$c$)$^{2}$.

In the plane-wave approximation, the recoil polarization produced by a
longitudinally polarized electron beam in quasielastic
electron-neutron scattering is restricted to the scattering plane
\cite{akhiezer74,arnold81}: The longitudinal component, $P_{L'}$, and
the transverse (sideways) component, $P_{S'}$, are parallel and
perpendicular, respectively, to the recoil neutron's momentum vector.
In terms of $G_{E}^{n}$ and $G_{M}^{n}$, $P_{S'}$ and $P_{L'}$ can be
written as
\setlength\arraycolsep{2pt}
\begin{eqnarray}
P_{S'}/P_{L} &=& -K_{S}G_{E}^{n}G_{M}^{n}/I_{0}~, \\
P_{L'}/P_{L} &=& \;\;\; K_{L}(G_{M}^{n})^{2}/I_{0}~,
\end{eqnarray}
where $P_{L}$ is the electron beam polarization,
$I_{0}{\equiv}(G_{E}^{n})^{2}+K_{0}(G_{M}^{n})^{2}$, and $K_{S}$, $K_{L}$,
and $K_{0}$ are kinematic functions of the electron scattering angle,
${\theta}_{e}$, and $Q^{2}$.  Measurements of $P_{S'}$ and $P_{L'}$
via a secondary analyzing reaction permit an extraction of the ratio
of $G_{E}^{n}$ to $G_{M}^{n}$; a significant advantage of this
technique is that $P_{L}$ and the analyzing power of the secondary
reaction cancel in the polarization ratio $P_{S'}/P_{L'}$. 
Also, for quasifree emission, Arenh\"{o}vel \cite{arenhoevel87} demonstrated that 
$P_{S'}$ and $P_{L'}$ are insensitive to final state interactions (FSI),
meson exchange currents (MEC), isobar configurations (IC), and to theoretical
models of deuteron structure.
In this Letter, we report
new measurements of $G_{E}^{n}/G_{M}^{n}$ obtained via recoil
polarimetry from the quasielastic $^{2}$H$(\vec{e},e'\vec{n})^{1}$H
reaction at three central $Q^{2}$ values of 0.45, 1.15, and 1.47
(GeV/$c$)$^{2}$.

\begin{figure}
\includegraphics[scale=0.70]{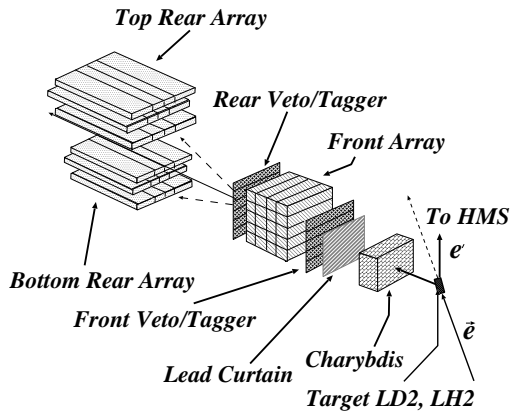}
\caption{\label{fig:figure1} A schematic diagram of the
polarimeter.}
\end{figure}

Our measurements were carried out in Hall C of the Thomas Jefferson
National Accelerator Facility.  
The experimental arrangement with an
isometric view of our polarimeter is shown in Fig.\
\ref{fig:figure1}.
A beam of longitudinally polarized electrons (with a typical
polarization of 80\%) 
scattered quasielastically from a neutron in a 15-cm liquid
deuterium target.  A scattered electron was detected in the High
Momentum Spectrometer (HMS) in coincidence with the recoil neutron.
The neutron polarimeter (NPOL) was used to measure the
up-down scattering asymmetry from the transverse component of the
recoil neutron polarization presented to the polarimeter.  
To permit measurements of
the up-down scattering asymmetry from different combinations of
$P_{S'}$ and $P_{L'}$, a dipole magnet (Charybdis) located in front of
the polarimeter precessed the recoil neutron's polarization vector
through an angle ${\chi}$.  

The polarimeter consisted of a total of 44 plastic scintillation
detectors.  To achieve luminosities of ${\sim}~3{\times}10^{38}$
cm$^{-2}$~s$^{-1}$, the front array was segmented into 20 detectors 
[100 cm$\times$10 cm$\times$10 cm].
Top and bottom rear arrays were
shielded from the direct path of particles from the target.
Each rear array consisted of 6 ``20-in'' detectors 
[101.6 cm$\times$50.8 cm$\times$10.16 cm]
and 6 ``10-in'' detectors [101.6 cm$\times$25.4 cm$\times$10.16 cm].
A double layer of ``veto/tagger'' detectors (each 0.64-cm thick) 
directly ahead of and
behind the front array identified incoming and scattered charged
particles.  A 10-cm lead curtain attenuated the flux of
electromagnetic radiation and charged particles incident on the
polarimeter.  The flight path from the center of the
target to the center of the front array was 7.0 m, and the mean flight
path from the front array to the rear array was 2.5 m.

For a fixed neutron scattering angle of 46.0$^{\circ}$, central $Q^{2}$
values of 0.45 and 1.47 (GeV/$c$)$^{2}$ were associated with beam
energies of 0.884 and 3.40 GeV, respectively, and electron scattering 
angles of 52.7$^{\circ}$ and 23.6$^{\circ}$, respectively.  
The measurement
conducted at a central $Q^{2}$ value of 1.15 (GeV/$c$)$^{2}$ was
associated with two beam energies of 2.33 and 2.42 GeV and electron
scattering angles of 30.8$^{\circ}$ and 30.1$^{\circ}$, respectively.  
We conducted
asymmetry measurements with the polarization vector precessed through
${\chi}={\pm}40^{\circ}$ at each of our $Q^{2}$ points; in addition, 
at $Q^{2}=1.15$ and 1.47 (GeV/$c$)$^{2}$, we conducted asymmetry
measurements with the polarization vector precessed through
${\chi}=0^{\circ},~{\pm}90^{\circ}$.  The acceptance-averaged values
of $Q^{2}$ are: ${\langle}Q^{2}{\rangle}=0.45$, 1.13, and 1.45
(GeV/$c$)$^{2}$.

\begin{figure}
\includegraphics[scale=0.80]{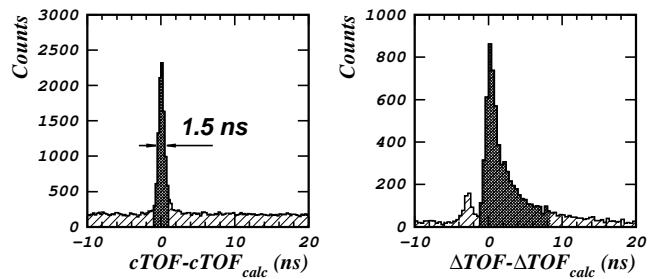}
\caption{\label{fig:figure2} Typical time-of-flight spectra
for $Q^2$ = 1.15 (GeV/$c$)$^{2}$.  The
dark-shaded regions indicate the selected portions of the spectra.}
\end{figure}

Typical time-of-flight spectra are shown in Fig.\
\ref{fig:figure2}.  The left panel is an HMS-NPOL coincidence
time-of-flight spectrum. We compared the measured time-of-flight, cTOF, with
the time-of-flight calculated from electron kinematics and
offsets determined by a calibration procedure; the result is centered
on zero with a FWHM of approximately 1.5 ns.  The right panel is the
time-of-flight spectrum between a neutron event in the front array and
an event in the top or bottom rear array.  We compared this measured
time-of-flight, ${\Delta}$TOF, with the time-of-flight
calculated for elastic $np$ scattering.  
Variations with respect to a 
nominal 2.5 m flight path were compensated.  
The tail on the slow side is due to Fermi motion in carbon and 
nuclear reactions, and the secondary
peak at ${\sim}-2.5$ ns is the result of ${\pi}^{0}$ production in the
front array.  To extract the physical scattering asymmetry, we
calculated the cross ratio, $r$, which is defined to be the ratio of
two geometric means, $(N_{U}^{+}N_{D}^{-})^{1/2}$ and
$(N_{U}^{-}N_{D}^{+})^{1/2}$, where $N_{U}^{+}(N_{D}^{-})$ is the
yield in the ${\Delta}$TOF peak for neutrons scattered up(down) when
the beam helicity was positive(negative); the yields, corrected for
background, were obtained by peak fitting.  The physical scattering
asymmetry is then given by $(r-1)/(r+1)$.
The merit of the cross ratio technique \cite{ohlsen73} is that the neutron
polarimeter results are independent of the luminosities for positive
and negative helicities, and the efficiencies and acceptances of the
top and bottom halves of the polarimeter.
Beam charge
asymmetries (of typically 0.1\%) and detector threshold differences
cancel in the cross ratio.

To account for the finite experimental acceptance and nuclear physics
effects such as FSI, MEC, and IC, we averaged
Arenh\"{o}vel's theoretical $^{2}$H$(\vec{e},e'\vec{n})^{1}$H
calculations \cite{arenhoevel95} over the experimental acceptance.
These calculations include leading-order relativistic contributions to
a non-relativistic model of the deuteron as an $n$-$p$ system, employ
the Bonn $R$-space $NN$ potential \cite{machleidt87} for the inclusion
of FSI, and include MEC and IC.  
Other realistic potentials (e.g., the Argonne V18 \cite{wiringa95}) give
essentially the same results. 
The theoretical values of the recoil
polarization were calculated over a kinematic grid; during the
averaging procedure, the recoil polarization was computed via
multidimensional interpolation between the grid elements.

To average these theoretical calculations over the
experimental acceptance, 
we prepared two independent simulation programs.  First, we developed
the \texttt{GENGEN} Monte Carlo simulation program, which
includes an event generator and detailed models of the electron
spectrometer and the neutron polarimeter.  \texttt{GENGEN} reproduces
experimental kinematic distributions and models the response of the
polarimeter.  Second, we developed a program that used the kinematics
of the reconstructed quasielastic events from the experimental data
to compute the recoil polarization for each event used in the
data analysis; the advantage of this method is that it does not require 
a model of the experimental acceptance.

For the first-pass analysis, the simulation programs used theoretical
calculations that assumed the Galster parameterization for $G_{E}^{n}$
with different multiplicative factors. We determined via simulation the 
optimal factor for each $Q^2$ that provided the best agreement between 
the simulated
polarization ratios and the experimental asymmetry ratios. Next we fitted
the current world data \cite{eden94,herberg99,
rohe99,golak01,passchier99,zhu01,schiavilla01,kopecky97} 
and our first-pass acceptance and
nuclear physics corrected results for $G_{E}^{n}$ to a Galster
parameterization with two free parameters.  Then we
repeated the simulations using new theoretical calculations that assumed
this modified Galster parameterization for $G_{E}^{n}$.  As in the
first-pass analysis, we determined
via simulation the optimal factor that provided the best agreement
between simulation and experiment.  The differences between these 
analyses were negligible, and the results from the two simulation
programs agreed to better than two percent.

The estimated values of the systematic uncertainties are listed in
Table~\ref{tab:systematics}.
A significant advantage of our experimental technique is that the
scale and systematic uncertainties are small; the analyzing power of
the polarimeter cancels in the polarization ratio, and the beam
polarization, $P_{L}$, also cancels as it varied little during
sequential measurements of the scattering asymmetries.  We
measured the beam polarization with a M{\o}ller polarimeter
\cite{hauger01}, and changes in $P_{L}$ were typically on the order of
one to two percent.  The helicity of the beam was reversed at a
frequency of 30 Hz to eliminate instrumental asymmetries.  

\begin{table}
\caption{\label{tab:systematics} Estimated systematic uncertainties
in ${\Delta}g/g$ $[$\%$]$.}
\begin{ruledtabular}
\begin{tabular}{lcccccc}
& \multicolumn{5}{c}{${\langle}Q^{2}{\rangle}$ $[$(GeV/$c$)$^{2}]$} \\
Source& 0.45\footnotemark[1]& 1.13\footnotemark[1]& 1.13\footnotemark[2]& 
1.45\footnotemark[1]& 1.45\footnotemark[2] \\ \hline
Beam Polarization&           1.4&   0.8& 0.4&  1.7& 0.3 \\
Charge Exchange&            $<$0.01&   0.02& 0.06&  $<$0.01& 0.2 \\
Depolarization&           $<$0.1& $<$0.1& 0.2&  0.1& 0.6 \\
Positioning/Traceback&       0.2&   0.3& 0.3&  0.4& 0.4 \\
Precession Angle&            1.1&   0.3& 0.1&  0.5& 0.1 \\
Radiative Corrections&       0.7&   0.1& 0.1& 0.05& 0.05 \\
\hline
Total of Above Sources&      1.9&   0.9& 0.5&  1.8& 0.8 \\
\end{tabular}
\end{ruledtabular}
\footnotetext[1]{${\chi}={\pm}40^{\circ}$ precession.}
\footnotetext[2]{${\chi}=0^{\circ},~{\pm}90^{\circ}$ precession.}
\end{table}

\begin{table*}
\caption{\label{tab:results} Results for $g=G_{E}^{n}/G_{M}^{n}$
and $G_{E}^{n}$.}
\begin{ruledtabular}
\begin{tabular}{cccc}
${\langle}Q^{2}{\rangle}$ $[$(GeV/$c$)$^{2}]$& $g=G_{E}^{n}/G_{M}^{n}$&
$G_{M}^{n}/{\mu}_{n}G_{D}$ \cite{kelly02}& $G_{E}^{n}$ \\ \hline
0.447& $-0.0761 \pm 0.0083 \pm 0.0021$& $1.003 \pm 0.006$&
$0.0550 \pm 0.0060 \pm 0.0016$ \\
1.132& $-0.131 \pm 0.010 \pm 0.003$& $1.057 \pm 0.017$&
$0.0394 \pm 0.0029 \pm 0.0012$ \\
1.450& $-0.190 \pm 0.016 \pm 0.004$& $1.044 \pm 0.024$&
$0.0411 \pm 0.0035 \pm 0.0013$ \\
\end{tabular}
\end{ruledtabular}
\end{table*}
A false asymmetry or a dilution of the asymmetry may arise from the
two-step process
$^{2}$H$(\vec{e},e'\vec{p})^{1}$H~+~Pb$(\vec{p},\vec{n})$; the
contamination from this process was assessed by running with a liquid
hydrogen target.  The contamination levels are negligible
(${\alt}$~0.3\%) for ${\chi}={\pm}40^{\circ}$ and ${\pm}90^{\circ}$
at all of our $Q^{2}$ points, and for ${\chi}=0^{\circ}$,
the contamination levels are ${\sim}$0.3\% and
${\sim}$3\% at ${\langle}Q^{2}{\rangle}=1.13$ and 1.45
(GeV/$c$)$^{2}$, respectively; accordingly, we have not corrected our
${\langle}Q^{2}{\rangle}=0.45$ and 1.13 (GeV/$c$)$^{2}$ data for
contamination from this two-step process.
The net correction obtained for the analysis of all of the data for  
${\langle}Q^{2}{\rangle}=1.45$ (GeV/$c$)$^{2}$ [viz., for
${\chi}=0^{\circ}$, ${\pm}40^{\circ}$, and ${\pm}90^{\circ}$] amounted
to 1.3$\pm$0.1\%.  In addition to
charge-exchange reactions in the lead curtain, the flux of neutrons
entering the polarimeter may be depolarized as a result of nuclear
interactions in the lead curtain.  Depolarization processes were
simulated in \texttt{GENGEN} using a spin-dependent
multiple-scattering algorithm employing quasifree scattering from a
Fermi gas.  The effects of depolarization cancel in the polarization
ratio, and the residual non-cancellation effect upon $g$ of less than 
0.6\% is included in the systematic uncertainty. 

\begin{figure}
\includegraphics[scale=0.55]{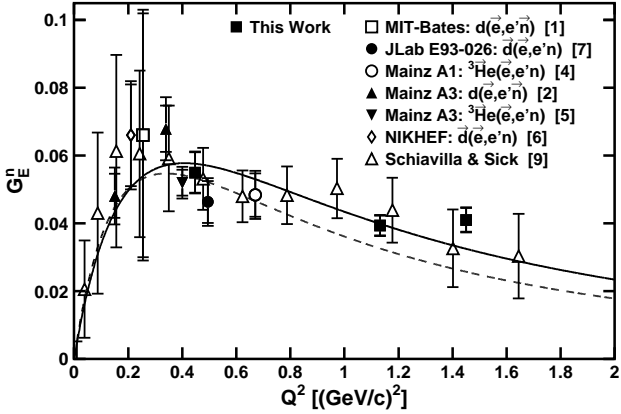}
\caption{\label{fig:figure3} The current world data on
$G_{E}^{n}$ versus $Q^{2}$ extracted from polarization measurements
and an analysis of the deuteron quadrupole form factor 
\cite{eden94,herberg99,rohe99,golak01,
passchier99,zhu01,schiavilla01}.  The Galster parameterization
\cite{galster71} is the dashed line, and our two-parameter Galster
fit (see text) is the solid line.}
\end{figure}
\begin{figure}[t]
\includegraphics[scale=0.55]{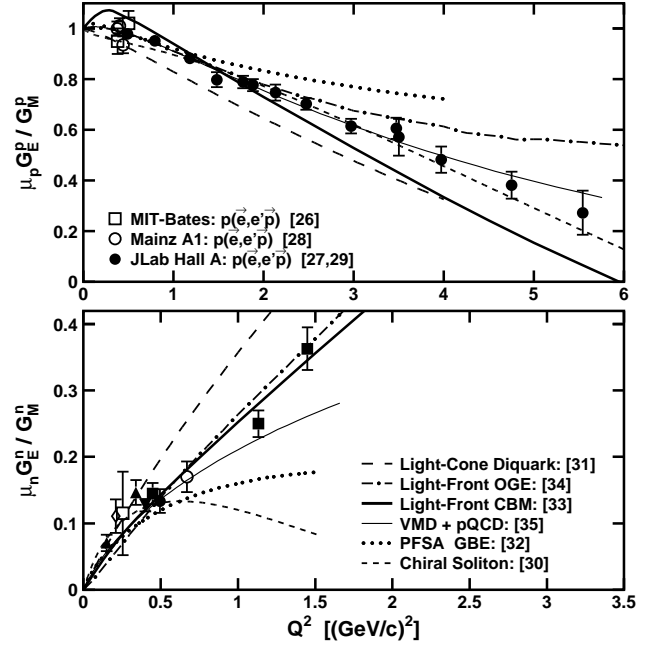}
\caption{\label{fig:figure4} Predictions of selected models (see text)
for ${\mu}_{p}G_{E}^{p}/G_{M}^{p}$ and
${\mu}_{n}G_{E}^{n}/G_{M}^{n}$ compared with proton
\cite{milbrath98,jones00,pospischil01,gayou02} and neutron
\cite{eden94,herberg99,rohe99,golak01,
passchier99,zhu01} data.  The neutron data symbols are the same as in
Fig.\ \ref{fig:figure3}.}
\end{figure}

Afanasev \textit{et al}.\ \cite{afanasev01} calculated radiative
corrections to the polarization-transfer coefficients, $P_{S'}/P_{L}$
and $P_{L'}/P_{L}$.  The primary effect is depolarization of the
electron such that both polarization-transfer coefficients should be
increased by ${\sim}$1.9\%, ${\sim}$3.7\%, and ${\sim}$4.4\% at
${\langle}Q^{2}{\rangle}=0.45$, 1.13, and 1.45 (GeV/$c$)$^{2}$,
respectively; however, these effects nearly cancel in the polarization ratio
such that the net effect upon $g$ is small at ${\langle}Q^{2}{\rangle}=0.45$ 
(GeV/$c$)$^{2}$ and negligible at the two higher $Q^2$ points.

The values of $g$ and $G_{E}^{n}$ that we report are listed in
Table \ref{tab:results}.  To determine our values for $G_{E}^{n}$, we used
the best-fit values for $G_{M}^{n}$ (listed in Table~\ref{tab:results}) 
obtained using the methods described in \cite{kelly02}. 
The new fit omits the data 
of \cite{markowitz93,bruins95} 
and includes the recent data of Xu \textit{et al}. \cite{xu02}.  The 
results are similar to those obtained by Friedrich and Walcher 
\cite{friedrich03} using a 
parameterization  designed to investigate the role of the pion cloud.  
The fit is also similar to that of Kubon \textit{et al}. \cite{kubon02} within 
their range of $Q^2$, but  extends to larger $Q^2$.
For $g$ and $G_{E}^{n}$, the first set of errors
is statistical, and the second set is systematic.
The quoted systematic uncertainties for $g$ include a 2\% uncertainty
that results when a different selection of runs is used for the time
calibration.
To obtain the systematic uncertainties for $G_{E}^{n}$,
the relative uncertainties in $G_{M}^{n}$ were added quadratically to the
systematic uncertainties listed in Table \ref{tab:systematics}
and the 2\% time calibration uncertainty.

Our values for $G_{E}^{n}$ are plotted in Fig.\
\ref{fig:figure3} together with the current world data on
$G_{E}^{n}$ \cite{eden94,herberg99,
rohe99,golak01,passchier99,zhu01,schiavilla01} extracted from
polarization measurements and an analysis of the deuteron quadrupole form
factor.  We fitted these data and the $G_{E}^{n}$ slope
at the origin as measured via low-energy neutron scattering 
from electrons in heavy atoms
 \cite{kopecky97} to a Galster parameterization: $G_{E}^{n} =
-a{\mu}_{n}{\tau}G_{D}/(1+b{\tau})$, where ${\tau}=Q^{2}/4M_{n}^{2}$,
$G_{D}=(1+Q^{2}/{\Lambda}^{2})^{-2}$, and ${\Lambda}^{2}=0.71$
(GeV/$c$)$^{2}$. Our best-fit parameters are
$a=0.888~{\pm}~0.023$ and $b=3.21~{\pm}~0.33$.

Polarization measurements of $G_{E}^{p}/G_{M}^{p}$ 
\cite{milbrath98,jones00,pospischil01,gayou02}
and $G_{E}^{n}/G_{M}^{n}$
\cite{eden94,herberg99,rohe99,golak01,
passchier99,zhu01,schiavilla01} are compared with predictions of
selected models in Fig. \ref{fig:figure4}.  The chiral soliton
model \cite{holzwarth02} reproduces the dramatic linear
decrease observed in ${\mu}_{p}G_{E}^{p}/G_{M}^{p}$ for $1<Q^{2}<6$
(GeV/$c$)$^{2}$; however, this model fails to reproduce the neutron
data at large $Q^{2}$.  
The light-cone diquark model \cite{ma02} achieves
qualitative agreement with the low $Q^{2}$ proton and neutron data;
however, at high $Q^{2}$, it lies below(above) the proton(neutron)
data.  A calculation using the point-form spectator approximation (PFSA)
with pointlike constituent quarks and a Goldstone boson exchange
interaction fitted to the meson and baryon spectrum 
\cite{boffi02} also achieves qualitative agreement with the low $Q^{2}$
proton and neutron data; however, it, too, fails to describe the high
$Q^{2}$ proton and neutron data.  A light-front calculation using
pointlike constituent quarks surrounded by a cloud of pions 
\cite{miller02} describes the neutron data, but falls below the
proton data at high $Q^{2}$.  A one-gluon exchange light-front
calculation using constituent quark form factors fitted to $Q^{2}<1$
(GeV/$c$)$^{2}$ data \cite{cardarelli00} agrees with the
neutron data, but deviates from the proton data above
$Q^{2}~{\sim}~3.5$ (GeV/$c$)$^{2}$.  Finally, fits that couple vector
meson dominance with the predictions of perturbative QCD 
\cite{lomon02} agree with the entire range of the proton data, but fall
below the neutron data above $Q^{2}~{\sim}~1.2$ (GeV/$c$)$^{2}$.

A successful model of confinement must be able to predict both
neutron and proton electromagnetic form factors simultaneously.  Although
these models achieve qualitative agreement with the proton data, the neutron
electric form factor is especially sensitive to small components of the
nucleon wave function and differences between model predictions for 
$G_{E}^{n}$ tend
to increase rapidly with $Q^2$.  Our new $G_{E}^{n}$ data for $0.5<Q^2<1.5$ 
(GeV/$c$)$^{2}$
provide a challenging test for confinement models and invite extensions to
higher $Q^2$.

We thank the TJNAF Hall C scientific and engineering staff for their
outstanding support.  Also, we thank A.\ Afanasev
for providing calculations of the radiative correction
uncertainties.  This work was supported in part by the National
Science Foundation, the U.S.\ Department of Energy, and the Deutsche
Forschungsgemeinschaft.  The Southeastern Universities Research
Association (SURA) operates the Thomas Jefferson National Accelerator
Facility under the U.S.\ Department of Energy contract
DE-AC05-84ER40150.


\end{document}